# Low Dimensional Material based Electro-Optic Phase Modulation Performance Analysis

Rubab Amin, Rishi Maiti, *Member, IEEE,* Jacob B. Khurgin, and Volker J. Sorger, *Senior Member, IEEE*

*Abstract*—Electro-optic modulators are utilized ubiquitously ranging from applications in data communication to photonic neural networks. While tremendous progress has been made over the years, efficient phase-shifting modulators are challenged with fundamental tradeoffs, such as voltage-length, index change-losses or energy-bandwidth, and no single solution available checks all boxes. While voltage-driven phase modulators, such as based on lithium niobate, offer low loss and high speed operation, their footprint of 10's of cm-scale is prohibitively large, especially for density-critical applications, for example in photonic neural networks. Ignoring modulators for quantum applications, where insertion loss is critical, here we distinguish between current versus voltage-driven modulators. We focus on the former, since current-based schemes of emerging thin electro-optical materials have shown unity-strong index modulation suitable for heterogeneous integration into foundry waveguides. Here, we provide an in-depth *ab-initio* analysis of obtainable modulator performance based on heterogeneously integrating low-dimensional materials, i.e. graphene, thin films of indium tin oxide, and transition metal dichalcogenide monolayers into a plurality of optical waveguide designs atop silicon photonics. Using the fundamental modulator tradeoff of energy-bandwidth-product as a design-quality quantifier, we show that a small modal cross section, such as given by plasmonic modes, enables high-performance operation, physically realized by arguments on charge-distribution and low electrical resistance. An in-depth design understanding of phase-modulator performance, beyond doped-junctions in silicon, offers opportunities for micrometer-compact yet energy-bandwidth-ratio constrained modulators with timely opportunities to hardware-accelerate applications beyond data communication towards photonic machine intelligence, for instance; where both performance and integration-density are critical.

## I. INTRODUCTION

THE ever-increasing growth of bandwidth demands technological breakthroughs of short reach interconnects [1], [2]. The performance of electrical interconnects is largely limited due to their loss, dispersion and electrical cross-talk thus optical interconnects are a potential solution for inter-chip but possible eventually also for in intra-chip distances [2], [3]. Silicon photonics offers a promising platform for future on-chip optical networks due to their low-cost electronic-photonic co-integration and availability of the matured CMOS technology [4], [5]. While silicon (Si) photonics is highly promising for optical routing, a complete optical interconnect solution requires other functionalities such as light generation, modulation, and detection, all are rather difficult to achieve in an entirely monolithic platform. Therefore, a hybrid heterogeneous integration of other emerging materials on Si photonic platform is key for future photonic integrated circuits [6], [7]. Recent advances on such heterogeneous integration of emerging low-dimensional materials showed promising properties such as their unit-strong index change capability and tunable bandgap, both with high relevance for on-chip compact and efficient photonic device applications. These emerging low dimensional materials include different material classes from transparent conducting oxide (TCO) thin films to more exotic solutions involving two dimensional (2D) materials such as graphene or transition metal dichalcogenide (TMD) nanocrystals, for instance. TCOs can offer the advantage of rapid foundry adaptability towards CMOS-near compatibility while 2D materials can provide additional advantage of freedom from lattice mismatch issues due to their out of plane Van der Waals bonding [8] – [15].

Here, in this work we focus our discussion on the electro-optic modulator, the workhorse-of-the-internet, converting high-speed electrical data into optical domain which underlies a plethora of applications ranging going beyond optical communications [16], but also for microwave and RF photonics [17], high-performance computing [18], and include facilitating emerging computing paradigms such as quantum and neuromorphic photonics [19], [20]. Depending on the optical property of the material (i.e. imaginary part or real part of the complex refractive index) variation for light modulation, electro-optic modulators can be categorized as electro-absorptive modulators and electro-refractive (phase) modulators. For electro-absorptive modulators, the absorption coefficient ($\kappa$) of the material is controlled by an absorption-related effect, such as saturable absorption, the Franz–Keldysh

The work of V. J. Sorger was supported in part by the Air Force Office of Scientific Research (AFOSR) under Grants FA9550-17-1-0071 and FA9550-17-1-0377. *(Corresponding author: Volker J. Sorger)*

Rubab Amin, Rishi Maiti and Volker J. Sorger are with the Department of Electrical and Computer Engineering, The George Washington University, Washington, DC 20052 USA (e-mail: rubabmn@gwu.edu; rmaiti11@gwu.edu; sorger@gwu.edu).

Jacob B. Khurgin is with the Electrical and Computer Engineering Department, Johns Hopkins University, Baltimore, MD 21218 USA (e-mail: jakek@jhu.edu).

effect, and the quantum-confined Stark effect [21], [22]. Whereas, electro-refractive modulators or phase modulators operate by changing the real part (*n*) of the complex refractive index traditionally by Pockel's or Kerr effect [23], [24]. However, it is impossible to modulate only real or imaginary parts independently as they are intertwined by a fundamental constraint, namely Kramers-Kronig (KK) relations. Hence, it is challenging to obtain a strong optical index change with the least amount of voltage (i.e. steepest switching), while observing optical loss limitations and ensuring micrometer compact device footprint at the same time. Conventional electro-optic materials such as silicon operating either with the plasma-dispersive carrier or Kerr effect show a rather low index change [25]. This necessitates heterogeneous integration of highly index tunable materials such as offered by low-dimensional materials discussed herein. Deploying these materials on bulk photonic modes can offer performance improvements despite their 'thin' dimension and subsequent reduced modal overlap factor [26]. Indeed, the use of plasmonic modes, for instance, can overcome the fundamental limit of footprint, speed, and power consumption by introducing the opportunity to shrink the active material from hundred's of nanometers (i.e. bulk modes) of Silicon or $LiNbO_3$ down to ten's of nanometers and enhance the light matter interaction (LMI) in the mode [27], [28]. As such, plasmonic modes do offer substantial performance improvements while costing insertion losses.

In addition, classifying based on physical mechanisms, another differentiation is the way the optical index change is induced, namely either being current-driven modulators versus voltage (field)--driven modulators. In the former, the index can be tuned by injecting (or depleting) charge carriers into (or from) the active region of the materials via capacitive gating [8], [9], [13] – [15], which is the focus of this discussion here. On the other hand, in case of voltage driven modulators, the index shift of the materials is caused by the strength of the electric field which essentially induces the energy level shift and/or the oscillator strength.

Here, we are interested in deriving a fundamental physical framework towards comparing disparate performance metrics and relevant tradeoffs for utilizing these emerging low dimensional materials in electro-optic integrated phase modulation. First, graphene is selected for its unique electrical and optical properties. Next, we choose indium tin oxide (ITO) as this is the most widely used TCO material in industry. Finally, we investigate tungsten di selenide ($WSe_2$) as a representative 2D material from the TMDs which have gained a renewed interest in the community from their bulk 3-dimentional counterparts. We investigate these materials for their phase-tuning capability corresponding to current driven schemes in an *ab-initio* approach, and explore their fundamental charge requirements to deliver a $\pi$ phase shift. We also appoint an array of modal structures comprising of both photonic and plasmonic options to extract relevant parameters for our *ab-initio* approach for these representative plurality of modes. Finally, we conduct a perturbative performance analysis for disparate active phase shifters relying on aforementioned modal structures to gain practical understanding of device performances and tradeoffs for electro-optic modulators.

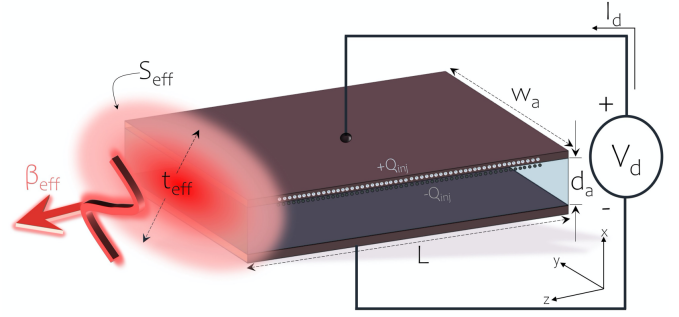

Fig. 1. Schematic representation of a generic current driven waveguide modulator. Drive voltage, $V_d$ causes a drive current $I_d$ to flow inducing charges $Q_{inj}$ on the gate/active layer interface. Light propagation constant inside the waveguide is shown with $\beta_{eff}$; active layer thickness is $d_a$; and the length and width of the modulator is represented with $w_a$ and $L$, respectively. The modal illumination pattern characterizes the effective area, $S_{eff}$ and effective thickness, $t_{eff}$ of the optical waveguide mode is shown. The corresponding coordinate system followed in the text is also depicted.

## II. PHASE MODULATOR ARRANGEMENT

First, we start off with a generic construction for our phase modulator study where we define relevant geometric waveguide parameters to our *ab-initio* approach. Then we define phase change cross-sections pertaining to the waveguide geometrical parameters and formulate necessary device metrics for the phase modulator analysis based upon those parameters.

### A. Effective Area and Thickness

A generic scheme exhibiting a waveguide integrated current-driven phase modulator showcases the modulator length, *L* and cross-sectional width, $w_a$ which is practically the waveguide width (Fig. 1). Both cladding layers on either side of the active layer (thickness, $d_a$) provide gating feasibility. Application of a drive voltage, $V_d$ facilitates a drive current, $I_d$ injecting charges into the active layer, $\pm Q_{inj}$ in the active layer (precisely put, at the interface of the actively tuned layer to its neighboring layer) formed by such electrostatic gating. Key to the modulator's performance is a strong LMI, which points towards requiring a small effective mode area, $S_{eff}$, or simply in one dimension (1D), a short effective thickness, $t_{eff}$.

Adopting the theory by Haus [29], shows the change of effective propagation constant of the waveguide as

$$\delta\beta = \omega \frac{\iint \delta\varepsilon(x,y)\mathbf{E} \cdot \mathbf{E}^* dxdy}{\iint \hat{z} \cdot \left(\mathbf{E} \times \mathbf{H}^* + \mathbf{E}^* \times \mathbf{H}\right) dxdy}$$
$$= \frac{\omega}{2c} \frac{\iint \delta\varepsilon_r(x,y)\mathbf{E} \cdot \mathbf{E}^* dxdy}{\frac{\eta_0}{2}\iint \hat{z} \cdot \left(\mathbf{E} \times \mathbf{H}^* + \mathbf{E}^* \times \mathbf{H}\right) dxdy} \quad (1)$$

Here, permittivity $\varepsilon = \varepsilon_0 \varepsilon_r$ and $\varepsilon_r$ is a tensor of dielectric constant, $\varepsilon_0$ and $\eta_0$ are the permittivity and impedance of free space, respectively.

Now, since we are dealing with the two dimensional active region, we may write $E(x,y) = E(x_0, y_0)e(x,y)$ where $e(x,y)$ is the normalized shape of the mode, $y_0$ is position of active layer so that $\delta\varepsilon_r(x,y) = \delta\varepsilon_{2D}\delta(y-y_0)$ $|x| \leq w_a/2$, where $w_a$ is the active layer width (we assume that $x=0$ in the middle of active layer). Therefore, after performing integration we obtain

$$\delta\beta = \frac{\omega}{2c} \frac{\delta\varepsilon_{2D} E^2(x_0, y_0) \int e_a^2(x,y_0)dx}{E^2(x,y_0) \int\int \hat{z}\cdot(e\times h^*)dxdy} = \frac{\omega}{2n_{eff}c} \frac{\delta\varepsilon_{2D}}{t_{eff}} \quad (2)$$

Here, $e_a$ is the projection of the normalized field onto the plane in which the dielectric constant is being changed, i.e. it is in-plane component of the field for 2D materials (e.g. graphene, TMD) and the total field for other bulk materials (e.g. ITO). We have introduced the dimensionless $h(x,y) = H(x,y)\eta_0/n_{eff}E_0$ and the effective thickness is

$$t_{eff} = \int\int \hat{z}\cdot(e\times h^*)dxdy / \int e_a^2(x,y_0)dx = S_{eff}/w_a \quad (3)$$

where the effective cross-section is

$$S_{eff} = w_a \int\int \hat{z}\cdot(e\times h^*)dxdy / \int e_a^2(x,y_0)dx \quad (4)$$

This definition is adaptable to the active layers that are few nm in thickness by simply averaging the field in them as shown in Appendix A. So that

$$\delta\beta = \frac{\omega}{2n_{eff}c} \frac{\delta\varepsilon_{2D} w_a}{S_{eff}} \quad (5)$$

We note, that (4) is indeed exact. The reason we have included $n_{eff}$ in definition of $h$ is that for a "weak waveguide" $H(x,y) \approx E(x,y)n_{eff}/\eta_0$ and they are normal to each other, so that $h \approx e$ and

$$S_{eff} \approx w_a \int\int e^2 dxdy / \int e_a^2(x,y_0)dx \quad (6)$$

which makes sense – if we assume that the waveguide is a rectangle $w\times t$ where the field is uniform and $w_a = w$ we find the area of rectangle. But the exact equation (4) actually incorporates the effect of group index – if the group index is large as in plasmonic waveguides, it means that the electric field in the metal is directed opposite to electric field in dielectric, and the integral in the numerator of (4) can become small.

III. INDEX MODULATION IN LOW DIMENSIONAL MATERIALS

We choose disparate emerging materials for integrated photonics to investigate phase modulation effects. These emerging low-dimensional materials have gained popularity in recent literature for their heterogenic integration compatibility with, for instance, the Si photonic platform. Here, we take an in-depth look into three modulation mechanism classes among these emerging novel materials including graphene (Pauli blocking and free carriers), indium tin oxide (ITO, free carrier dispersion), and WSe$_2$ (exciton modulation via carriers).

A. Graphene

Graphene as a phase modulator bears a unique feature – two different mechanisms of index change acting in unison when more carriers get injected. The first mechanism is associated with the reduction of refractive index as the Fermi energy $E_F$, and thus, the interband absorption edge moves towards higher energies. The second mechanism is reduction of index due to increase of (negative) free carrier response in accordance with the Drude formula. At the same time, the upshift of absorption edge causes decrease of interband absorption, but an increase in the number of free carriers causes the opposite effect. As a result, it is possible to select operating conditions in such a way that index modulation is strong while the absorption modulation is fully cancelled. Such arrangement would be highly desirable as it would facilitate to achieve full zero transmission in Mach-Zehnder modulators, and would reduce nonlinear distortion. To find out what are the conditions for such pure phase modulation, we consider the absorption coefficient due to interband transitions of a single layer graphene in a waveguide, which can be written as (Appendix B)

$$\alpha_{ib} = \frac{w_a}{n_{eff}S_{eff}}\pi\alpha_0 \frac{1}{e^{\frac{E_F - \hbar\omega/2}{kT}} + 1} \quad (7)$$

where $\alpha_0$ is a fine structure constant. Change of the absorption with the change of Fermi energy (Pauli blocking) is

$$\frac{d\alpha_{ib}}{dE_F} = -\frac{w_a}{n_{eff}S_{eff}}\pi\alpha_0 \frac{1}{4kT\cosh^2\left(\frac{E_F - \hbar\omega/2}{2kT}\right)} \quad (8)$$

Absorption due to free carriers is (see appendix B)

$$\alpha_{fc} = \frac{w_a}{n_{eff}S_{eff}}\alpha_0 \frac{4\gamma E_F}{\hbar(\omega^2 + \gamma^2)} \quad (9)$$

where $\gamma$ is the momentum scattering rate of the carriers. And obviously

$$\frac{d\alpha_{fc}}{dE_F} = \frac{w_a}{n_{eff}S_{eff}}\alpha_0 \frac{4\gamma}{\hbar(\omega^2 + \gamma^2)} \quad (10)$$

So, the total change is

$$\frac{d\alpha}{dE_F} = \frac{w_a}{n_{eff}S_{eff}}\pi\alpha_0 \left[\frac{4}{\pi\hbar\omega}\frac{\gamma}{\omega} - \frac{1}{4kT\cosh^2\left(\frac{E_F - \hbar\omega/2}{2kT}\right)}\right] \quad (11)$$

Here, we have neglected $\gamma^2$ in the denominator, because $\gamma \sim 2THz \ll \omega = 2\pi\times 200THz$ [30]. Absorption and its change are plotted in Fig. 2(a) and 2(b), and a closer look at Fig.2(c) reveals that at one particular photon energy the change of absorption becomes zero. This allows us to find the exact value of the position of Fermi level at which absorption remains constant at a given wavelength ($\lambda = 1550nm$) as the Fermi energy changes. It is easy to see that away from Fermi level cosh can be approximated by the exponential, and one obtains

$$E_{F0} - \hbar\omega/2 \approx kT\log\frac{\pi}{4}\frac{\hbar\omega}{kT}\frac{\omega}{\gamma} \approx 10kT = 250meV \quad (11a)$$

As such, for telecom C band operations ($\hbar\omega = 800meV$), we find, with the Fermi level at $E_{F0} = 650meV$ a required electron density of $n_0 = E_{F0}^2 / \pi \hbar^2 v_F^2 \sim 3 \times 10^{13} \, cm^2$, which is quite high. However, such densities are attainable and have been reported in recent literature [31], [32]. We can estimate the insertion loss at this value of $n_0$ by substituting condition (11a) into (7) and (9) and summing them up

$$\alpha = \frac{4w_a}{n_{eff} S_{eff}} \alpha_0 \frac{\gamma(E_{F0} + kT)}{\hbar \omega^2} \approx 0.005 \alpha_0 \frac{w_a}{n_{eff} S_{eff}} \quad (12)$$

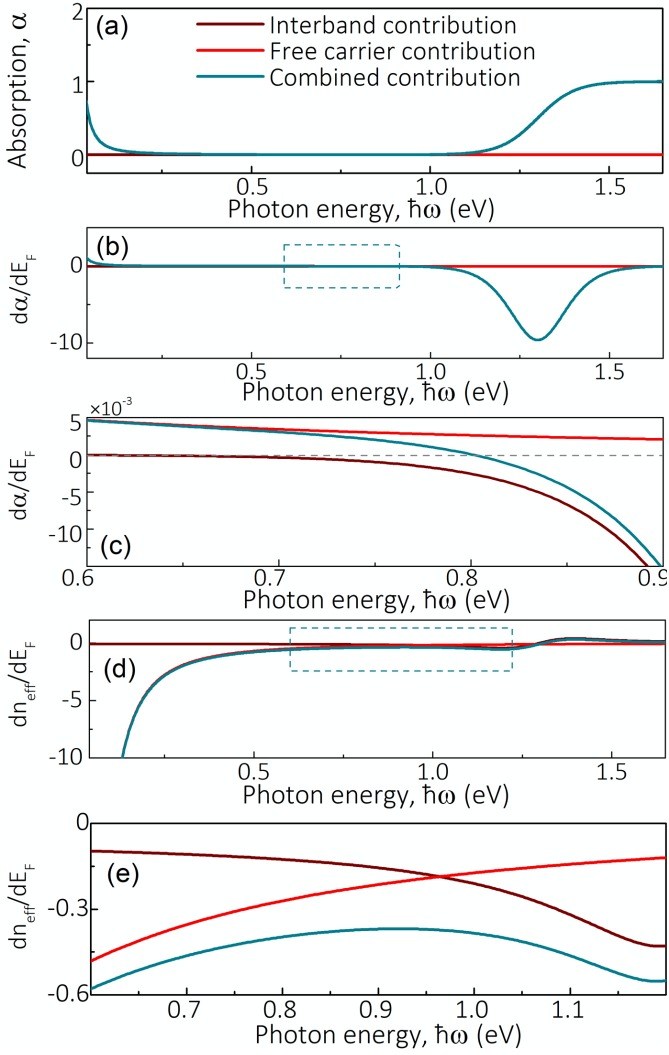

Fig. 2. (a) Graphene absorption, $\alpha$ in units of $w_a \pi \alpha_0 / n_{eff} S_{eff}$; (b) Change of absorption, $d\alpha/dE_F$ in units of $w_a \pi \alpha_0 / n_{eff} S_{eff}$ $(1/eV)$; (c) Magnified view of panel (b) dashed box area; (d) Change of effective index, $dn_{eff}/dE_F$ in units of $w_a c \alpha_0 / n_{eff} S_{eff}$ $(1/eV)$; and (e) Magnified view of panel (d) dashed box area. All parameters in (a) – (e) are plotted vs. photon energy, $\hbar\omega$ in unites of eV. All relevant dimensions are in cm.

The first term in parenthesis being free carrier absorption and the second, much smaller one is interband absorption

Next, we calculate the change of the refractive index using usual KK relation as

$$\delta n_{eff} = \frac{c}{\pi} \int_0^\infty \frac{\delta \alpha(\omega_1)}{\omega_1^2 - \omega^2} d\omega_1 \quad (13)$$

Now, the function $1/4kT \cosh^2(E_F - \hbar\omega/2/2kT)$ has a full width half maximum (FWHM) ~ $8kT$, thus when we are calculating the obtainable index change (real-part) at a detuning of, say, $10kT$ we can approximate it as $2\delta(\omega - 2\hbar^{-1}E_F)/\hbar$ and obtain for the index change due to interband transition as

$$\frac{dn_{ib}}{dE_F} = -\frac{w_a}{n_{eff} S_{eff}} \alpha_0 \frac{2c}{\hbar} \frac{1}{4E_F^2/\hbar^2 - \omega^2}$$
$$= -\frac{w_a}{n_{eff} S_{eff}} \frac{\alpha_0}{\hbar \omega^2} \frac{2c}{(2E_F/\hbar\omega + 1)(2E_F/\hbar\omega - 1)} \quad (14)$$

For the free carrier induced shift we obtain

$$\frac{dn_{fc}}{dE_F} = -\frac{w_a}{n_{eff} S_{eff}} \alpha_0 \frac{2c}{\hbar(\omega^2 + \gamma^2)} \approx -\frac{w_a}{n_{eff} S_{eff}} \frac{\alpha_0 \lambda_0}{\pi \hbar \omega} \quad (15)$$

And so the total effective index change plotted in Fig. 2(d) and (e) is

$$\frac{dn_{eff}}{dE_F} = -\frac{w_a}{n_{eff} S_{eff}} \frac{\alpha_0 \lambda_0}{\pi \hbar \omega} \left[ 1 + \frac{1}{(2E_F/\hbar\omega)^2 - 1} \right]$$
$$\approx -f \frac{w_a}{n_{eff} S_{eff}} \frac{\alpha_0 \lambda_0}{\pi \hbar \omega} \quad (16)$$

where $f \sim 1.6$ near $\hbar\omega = 0.8 eV$. The absorption, effective index and relevant changes in them with variations in the Fermi level are shown in Fig. 2. Note, the modulating absorption becomes null near the operating energy levels for aforementioned parameters (Fig. 2(c)). Corresponding effective index change for variations in the Fermi level are also shown (Fig. 2(d, e)). Now we need to find the expression for the change of optical index with the change of sheet (2D) carrier density. Using the relation

$$N_{2D} = E_F^2 / \pi \hbar^2 v_F^2; \quad \frac{dN_{2D}}{dE_F} = \frac{2E_F}{\pi \hbar^2 v_F^2} = \frac{2N_{2D}}{E_F} \quad (17)$$

we obtain

$$\frac{dn_{eff}}{dN_{2D}} = -\frac{w_a}{n_{eff} S_{eff}} f \frac{\alpha_0 \lambda_0}{\pi \hbar \omega} \frac{E_F}{2N_{2D}} \quad (18)$$

Then we can evaluate the phase shift as

$$\delta \Phi = \frac{2\pi}{\lambda_0} L \frac{dn_{eff}}{dN_{2D}} \delta N_{2D}$$
$$= \frac{2\pi}{\lambda_0} \frac{w_a L \delta N_{2D}}{n_{eff} S_{eff}} f \frac{\alpha_0 \lambda_0}{\pi \hbar \omega} \frac{E_F}{2N_{2D}} \delta n = \frac{\delta Q}{e n_{eff} S_{eff}} \sigma_\Phi \quad (19)$$

where $\delta Q = e w_a L \delta n$ is the injected (generated) charge and

$$\sigma_\Phi = f \frac{\alpha_0}{N_{2D}} \frac{E_F}{\hbar \omega} = f \alpha_0 \pi \frac{\hbar^2 v_F^2}{\hbar \omega E_F} \approx 3 \times 10^{-16} \, cm^2 \quad (20)$$

is the phase change cross section – change of phase per one injected carrier per unit cross-section. For full switching 180º ($\pi$) phase shift is required, therefore the switching charge can be evaluated as

$$Q_\pi = \pi \frac{e n_{eff} S_{eff}}{\sigma_\Phi} \quad (21)$$

As one can see, the switching charge, and hence switching energy depends only on two cross-sections – one is a material parameter $\sigma_\Phi$ and the other is a waveguide parameter $S_{eff}$. This, then, suggests studying the material and waveguide variations separately and explore different combinations of them.

Of course, one can also introduce half-wave voltage as $V_\pi = Q_\pi / C$ or

$$V_\pi L = Q_\pi t_{ox} / \varepsilon_0 \varepsilon_{ox} w_{eff} = \frac{e n_{eff} t_{ox} S_{eff}}{\varepsilon_0 \varepsilon_{ox} w_{eff} \sigma_\Phi} \quad (22)$$

where $t_{ox}$ and $\varepsilon_{ox}$ are the thickness and relative permittivity of the gate dielectric, respectively; and $w_{eff}$ is the effective width of the structure. However, this expression includes the gate dielectric thickness and permittivity, which means that it is not a proper measure of how effective a given material is intrinsically for phase modulation.

The change in the Fermi energy required to obtain a $\pi$ phase shift can be found from

$$2 f \frac{w_a L}{n_{eff} S_{eff}} \frac{\alpha_0}{\hbar \omega} \delta E_F = \pi \quad (23)$$

And substituting this into (12), the insertion loss is

$$\alpha L = \frac{2\pi}{f} \frac{\gamma}{\omega} \frac{E_{F0} + kT}{\delta E_F} \approx \frac{12}{\delta E_F} \quad (24)$$

where $\delta E_F$ is in units of meV. Therefore, the insertion loss of graphene modulator can be held down to about 1 dB if $\delta E_F \sim 50 meV$ and scattering rate $\gamma$ is aforementioned 2 THz.

### B. Indium Tin Oxide (ITO)

For ITO, one can obtain the index change simply via KK relations. The index change is (see appendix C)

$$\frac{dn_{eff}}{dN_{2D}} = -\frac{w_a}{2 n_{eff} S_{eff}} \frac{e^2}{\varepsilon_0 m^* (\omega^2 + \gamma^2)} \quad (25)$$

where $m^*$ is the optical effective mass of the conduction band. The corresponding phase change is then

$$\delta \Phi = \frac{\omega}{c} L \frac{dn_{eff}}{dN_{2D}} \delta N_{2D}$$
$$= -\frac{w_a L \delta N_{2D}}{2 n_{eff} S_{eff}} \frac{\omega e^2 / \varepsilon_0 c m^*}{\omega^2 + \gamma^2} = -\frac{\delta Q}{e n_{eff} S_{eff}} \sigma_\Phi \quad (26)$$

The phase change cross-section is (neglecting $\gamma$)

$$\sigma_\Phi = \frac{\omega e^2 / 2 \varepsilon_0 c m^*}{\omega^2 + \gamma^2} \approx 2\pi \alpha_0 \frac{\hbar^2}{m_0} \frac{m_0}{m^*} \frac{1}{\hbar \omega} \approx 10^{-16} cm^2 \quad (27)$$

And the $\pi-$ switching charge is still described by (21). With the absorption change being

$$\delta \alpha L = (2\gamma/c) \delta n_{eff} L = 2\pi \gamma / \omega \quad (28)$$

This change can be kept within 1 dB but obviously cannot be decoupled from the index change. The overall insertion loss is then

$$\alpha L = \alpha_{ON} L + 2\pi \gamma / \omega \quad (29)$$

where $\alpha_{ON}$ is the corresponding modal absorption due to the lossy ITO material at the high transmission state during modulation. The scattering rate, $\gamma = \hbar / \tau$ where $\tau$ is the scattering time which typically ranges around a few fs [33].

### C. Tungsten di-Selenide (WSe$_2$)

Transition metal dichalcogenides (TMDs) are a novel class of 2D materials that are gaining renewed interest in the community for their unique properties, such as tunable bandgap as a function of strain [12], for instance. Optical modulation manipulating the strong excitonic resonance of the monolayer TMDs suggests promising modulation performances. Here we chose WSe$_2$ as a representative TMD material whereas this modulation mechanism and the perturbative performance analysis can be extended to other similar TMD materials also. The change of the real part of the effective index with 2D carrier concentration change is caused by two phenomena – exciton screening and phase space filling [34], [35], causing the reduction of oscillator strength and can be expressed with the help of KK relations as

$$\frac{dn_{eff}}{dN_{2D}} = -\frac{w_a}{2 n_{eff} S_{eff}} \frac{e^2 P^2 / m_0^2 \omega^2 (\omega_0 - \omega)}{2 \varepsilon_0 \hbar \left[(\omega_0 - \omega)^2 + \gamma^2\right]}$$
$$\approx -\frac{w_a}{2 n_{eff} S_{eff}} \frac{e^2}{4 \varepsilon_0 m^* \omega^2 (\omega_0 - \omega)} \quad (30)$$

where $\omega_0$ is the exciton resonance frequency. Here, a factor of 2 in denominator comes from the fact that the oscillator strength is split between two polarizations, and the relation between momentum matrix element $P$ for the valence-to-conduction band transition, and the effective mass of the conduction band [36]

$$\frac{m_0}{m^*} = 1 + \frac{2P^2}{m_0 E_{gap}} \approx \frac{2P^2}{m_0 \hbar \omega} \quad (31)$$

has been used. If we compare this with ITO we can see that $4(\omega_0 - \omega)$ in denominator of (34a) is not far from $\omega$ in (28) so the difference in strengths of EO effect is simply related to the electron ability to move, quantified by the (inverse) effective mass.

We can also repeat all the calculations for ITO and find the phase change cross section as

$$\sigma_\Phi = \frac{-e^2 / 4 \varepsilon_0 c m^* (\omega_0 - \omega)}{(\omega_0 - \omega)^2 + \gamma^2}$$
$$\approx \pi \alpha_0 \frac{\hbar^2}{m_0} \frac{m_0}{m^*} \frac{1}{\hbar(\omega_0 - \omega)} \approx 10^{-16} cm^2 \quad (32)$$

The FWHM of exciton transitions typically are in the order of a few meV [37], as such, we have chosen the detuning $\omega_0 - \omega = 3\gamma$ and the effective mass, $m^* \sim 0.2 m_0$ is used. It should be noted that explicitly this expression does not contain any trace of the excitonic nature of the transition. As discussed in our previous work [38], the stronger excitonic transition is, the more robust it is against changes imposed by the injected carriers. It is difficult to screen strong exciton and in the end what matters is that addition of each electron reduces the total oscillator strength of an exciton by the strength of a single conduction – to – valence state absorption. A major difference between ITO and TMD, is that for TMD, one has an additional variable parameter – detuning $\omega_0 - \omega$. It is not difficult to obtain the expression for the change of insertion loss, similar to (28)

$$\delta\alpha L = 2\pi\gamma / (\omega_0 - \omega) \qquad (33)$$

Therefore, by reducing the detuning from the exciton transition it may be possible to enhance the phase change cross-section and thus reduce switching charge and energy, but only by paying the price of increased absorption modulation.

## IV. MODAL PARAMETERS FOR DIFFERENT WAVEGUIDES

We choose different realistic modes for our relevant parameter extraction combining both photonic and plasmonic designs. Our choice of the three different low-dimensional active materials coupled with photonic/plasmonic configurations puts our waveguide consideration array into 6 different modal structures. Every waveguide based modulator modal structure in this study are chosen on the Si on insulator (SOI) platform with $SiO_2$ substrate. All the photonic modal structures are designed with a widely used Si photonic waveguide dimensions of 500 nm × 220 nm, constrained by the epi-Si height of standard foundry processes and diffraction limit of C-band NIR light in Si. The photonic modal structures place the low-dimensional active material over the Si waveguide separated by a thin film of dielectric (oxide) facilitating electrostatic gating schemes (Fig. 3). The plasmonic modes are chosen with a view to maximize modal confinement and enhance attainable modulator performance aided by experimental demonstrations in recent years [8], [9], [39]. The ITO plasmonic mode has been chosen in a hybrid plasmon polritonic (HPP) configuration [8], [40], [41] with a transverse magnetic (TM)-like polarization to capture more light into the active material and enhance modulation metrics. On the other hand, graphene and $WSe_2$ plasmonic modes are chosen in the slot configuration as this structure has been shown to enhance in-plane electric field interaction with the monolayer active materials in recent literature [39] – [41]. The transverse electric (TE)-like polarization facilitated by the slot modal structure helps to interact skillfully with the monolayer 2-dimensional graphene and $WSe_2$ flakes, as the selective electric field in such case is in-plane with the flake. All gate oxide thickness are kept constant across all the modes with $t_{ox}$ = 10 nm, and same relatively high-$k$ dielectric $Al_2O_3$ is chosen. All the metal utilized in the plasmonic structures are Au and the thickness is also kept constant, $t_{Au}$ = 50 nm for a levelled comparison. The effective modal area, $S_{eff}$ (Fig. 3(c)) for these realistic modes with active layer thicknesses are calculated according to Appendix A.

## V. PHASE MODULATOR PERFORMANCE

As expected from our prior work on electro absorption modulators [38], the most relevant material parameter for phase modulation, the phase change cross-section $\sigma_\Phi$ shown in Table I does not vary much among all the materials considered here. That is of course easy to understand as the index change is the change of total polarizability caused by injection (depletion) of carriers. The easier it is for carriers to move and the closer is the frequency to resonance, the larger is the polarizability. Since it is paramount to avoid absorption, the detuning from the resonance is on the order of the operating frequency for all cases. Ability of motion is defined by the inverse effective mass (or, in the case of graphene by $v_F^2 / 2E_F$) which gives some advantage to graphene in which $\sigma_\Phi$ is almost three times higher (Table I). Another important advantage of graphene, as mentioned above, is the fact that it can provide (at least to the first order) pure phase modulation without spurious amplitude modulation. But these advantages can be negated by the fact that a monolayer graphene does not possess high enough density of states in which the injected carriers can reside.

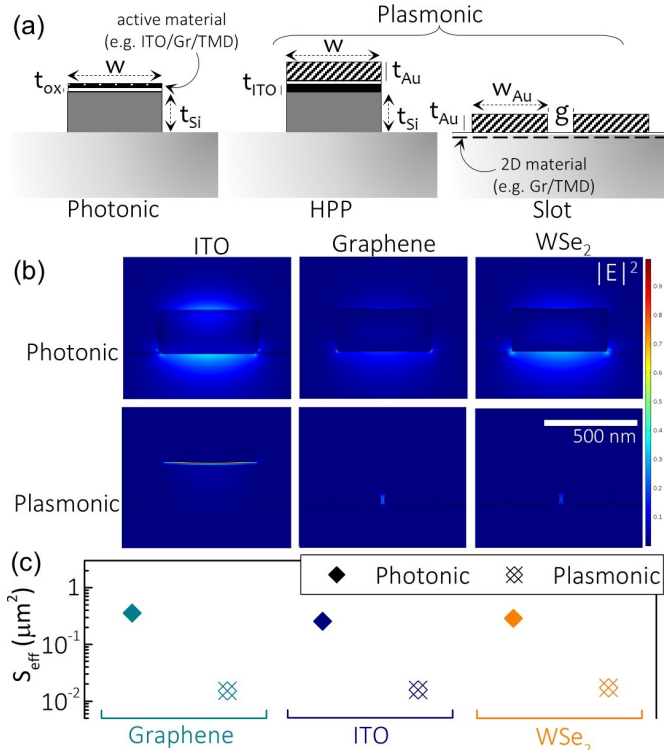

Fig. 3. (a) Schematic cross-sectional views of different modal configurations combining photonic and plasmonic designs. Relevant parameters are: $w$ = 500 nm, $t_{Si}$ = 220 nm, $t_{ox}$ = 10 nm, $t_{ITO}$ = 10 nm, $t_{Au}$ = 50 nm, $w_{Au}$ = 300 nm, and $g$ = 20 nm. Monolayer graphene and $WSe_2$ thickness are taken as 0.35 nm and 0.7 nm, respectively; (b) Mode profiles of the selected modal structures obtained with FEM eigenmode analysis, normalized $|\mathbf{E}|^2$ is plotted across the different modes where the color bar ranges from 0 to 1. (c) Effective modal area, $S_{eff}$ in $\mu m^2$ are extracted from the FEM eigenmode analyses for all photonic/plasmonic modes. (See sec. IIA for definitions).

TABLE I
PHASE CHANGE CROSS-SECTIONS, $\sigma_\Phi$ COMPARISON

| Material | Expression | Approximate Results |
|---|---|---|
| Graphene | $\sigma_\Phi = f\alpha_0 \pi \dfrac{\hbar^2 v_F^2}{\hbar\omega E_F}$ | $3 \times 10^{-16}\, cm^2$ |
| ITO | $\sigma_\Phi = 2\pi\alpha_0 \dfrac{\hbar^2}{m^*} \dfrac{1}{\hbar\omega}$ | $10^{-16}\, cm^2$ |
| $WSe_2$ | $\sigma_\Phi = \pi\alpha_0 \dfrac{\hbar^2}{m^*} \dfrac{1}{\hbar(\omega_0 - \omega)}$ | $10^{-16}\, cm^2$ |

Building upon the analytical material dependent results in sec. III, combined with realistically attainable modal parameters with our chosen modes in sec. IV, in this section we verify these perturbative results to reflect actual device performance (Fig. 4). First, we calculate the change of effective modal index, $\Delta n_{eff}$ based on the injected carriers on the gate dielectric and active material surface, $\Delta N_{2D}$. For discrete modulation states, the effective index change in graphene can be approximated as

$$\Delta n_{eff} = -\frac{w_a}{n_{eff} S_{eff}} f \frac{\alpha_0 \lambda_0}{\pi \hbar \omega} \frac{E_{F0}}{2n} \Delta N_{2D} \quad (34)$$

where we use the Fermi level as 650 meV as shown in sec. IIIA and (11), and corresponding carrier level $\sim 3 \times 10^{13}$ cm$^{-2}$. Similarly, for the ITO case we use,

$$\Delta n_{eff} = -\frac{w_a}{2 n_{eff} S_{eff}} \frac{m_0 e^2}{\varepsilon_0 m^* (\hbar^2 \omega^2 + \hbar^2 \gamma^2)} \frac{\hbar^2}{m_0} \Delta N_{2D} \quad (35)$$

where $\hbar^2 / m_0 = 7.62 eV \cdot A^2$ and $e^2 / \varepsilon_0 = 181 eV \cdot A$. $\gamma = \hbar / \tau$ is the collision frequency of the carriers in the ITO, and $\tau$ is the scattering time. We choose a scattering time of 5 fs consistent with our previous experimental results of ITO material's metrology and spectroscopic ellipsometry [8], [33]. The effective mass is taken as $m^* = 0.35 m_0$ [40] – [43].

Finally, the carrier concentration change dependency of the effective index change for WSe$_2$ is taken as

$$\Delta n_{eff} = -\frac{w_a}{2 n_{eff} S_{eff}} \frac{m_0 e^2}{4 \varepsilon_0 m^* \hbar \omega \hbar (\omega_0 - \omega)} \frac{\hbar^2}{m_0} \Delta N_{2D} \quad (36)$$

where a detuning of $\hbar(\omega_0 - \omega) = 0.2 eV$ is chosen as operational region and the effective mass is taken as $m^* = 0.25 m_0$ [38]. We plot the effective indices change, $\Delta n_{eff}$ as a function of the injected carriers, $\Delta N_{2D}$ induced by electrostatic gating in Fig. 4(a) for all the chosen modes in sec. IV. Note, $w_a = w$ for all photonic modes and the ITO HPP mode in (34) – (36) and Fig. 3(a); and $w_a = 2 w_{Au} + g$ for the plasmonic slot modes utilized for the graphene and WSe$_2$ plasmonic cases.

The results (Fig. 4) are what was expected by the analysis done in section III that indicate that graphene has an advantage over the other materials, i.e. ITO and TMD, which can be traced to the fact that the equivalent effective mass of graphene $m_{eq}^* / m_0 = (2 E_F / f m_0 c^2)(c / v_F)^2 \approx 0.14$ is small. Also, the plasmonic modulators show greatly enhanced performance, but, obviously at the expense of the increased insertion loss.

Next, we calculate the active device lengths required for $\pi$ phase shifts for all different modal structures based on the maximum (and minimum) attainable effective index change for our carrier concentration conditions (Fig. 4(e)). We note that, the carrier concentration can reach about one order of magnitude more in ITO than is portrayed in Fig. 4(a); but we opted to compare all different active materials on levelled circumstances and the linear dependency does not hold true for Pauli blocking based schemes such as in graphene and WSe$_2$ for such high carrier levels. Also, in order to switch the higher carrier levels, voltage requirements also would rise necessitating the use of higher-k gate dielectrics such as HfO$_2$, HfN, ZrN, etc.

Additionally, we evaluate the attainable phase shift, $\Delta \Phi = (2\pi / \lambda_0) \Delta n_{eff} L$ with the amount of injected charge using $Q_{inj} = e \Delta N_{2d} w_a L$ (Fig. 4(b)). The gate capacitance can be estimated using $C_g = \varepsilon_0 \varepsilon_{ox} (w_a L / t_{ox})$ for the active phase shifters with length dependency (Fig. 4(f)). We use Al$_2$O$_3$ as the gate oxide and the thickness is kept constant at $t_{ox} = 10 nm$ across all the different modal structures in order to calibrate a unifying performance metric for all active phase shifters. We further obtain the bias voltage requirements for the modulation as $V_d = Q_{inj} / C_g = (e / \varepsilon_0 \varepsilon_{ox}) t_{ox} N_{2D}$ and determine the voltage required for $\pi$ phase shifts, $V_\pi$ from there for both lengths for the maximum and minimum carrier sweep levels (Fig. 4(c); left dotted line for the minimum carrier sweep and the right solid line for maximum carrier sweep). As expected, the $V_\pi$ scales inversely in a linear manner with the lengths of the devices. Note, for any given carrier concentration change and corresponding effective index change, the dimensions of the active devices do not play a role in determining the bias voltage as the device dimensions cancel out; hence, the voltage requirements remain same across different modal structures. Next, the dynamic switching energy per bit is estimated as $U_{sw} = \frac{1}{4} C_g V_d^2 = \frac{1}{4} Q_{inj}^2 / C_g^2$ for the different modal phase modulator operation (Fig.4(d)). It should be mentioned for clarity that both the phase shift, $\Delta \Phi$ and dynamic energy, $U_{sw}$ shown (Fig. 4(b) and 4(d)) correspond to the maximum carrier sweep facilitated maximum effective index change (Fig. 4(a)) and corresponding minimum modulator lengths, $L_\pi$ (Fig. 4(e)) as showing similar results for the minimum carrier sweeps in the same panel would only clutter the discernable information.

Consequently, we evaluate the energy per bit needed for $\pi$ phase shift as $U_\pi = \frac{1}{2} C_g V_\pi^2$ for the length (i.e. carrier concentration change) dependent sweep also (Fig. 4(g)). We further calculate the -3 dB cutoff frequency of the different modulators as $f_{-3dB} = 1 / 2\pi R C_g$ for the carrier concentration change dependent length sweep (Fig. 4(h)). We have taken a nominal $R = 50 \Omega$ in these calculations, in order to show fundamental modulation bandwidth potential for these schemes. However, such a low resistance may not be realistic for the photonic modes, where partial and selective doping has to be used in order to keep both the carrier density, and hence, optical loss low [38], or the selected TMD and possibly ITO materials. In contrast, the metallic layer deployed in plasmonics can aid to serve as a low resistance contact as a byproduct allowing to define the electrical capacitor with high spatial overlap to the actual device region [44]. The switching energy per bit and attainable bandwidth (speed) of the active modulator are key performance metrics and scale inversely to each other with the gate dielectric thickness and material choice (i.e. screening), in principle. Accordingly, we previously introduce a more relevant figure of merit for modulators – the ratio of switching energy to the -3 dB bandwidth as energy bandwidth ratio (EBR) [38]. The goal for highly performing devices is to aim to minimize the numerical value for this performance metric. Here, we find this figure of merit as

$$EBR = \pi R C_g^2 V_\pi^2 = \pi R Q_\pi^2 \qquad (37)$$

where $Q_\pi$ is the corresponding charge required at the gate for $\pi$ phase shift in modulation. The EBRs for all the phase shifters are calculated, and evidently, the length (carrier concentration change) does not influence this metric (Fig. 4(i)). The underlying charge for these current driven modulators corresponding to $\pi$ phase shifts are also determined (Fig. 4(j)). The carrier concentration change and corresponding active device length variations also do not influence the $Q_\pi$ as the fundamental charge requirement to alter the optical phase remains the same in a given waveguide mode.

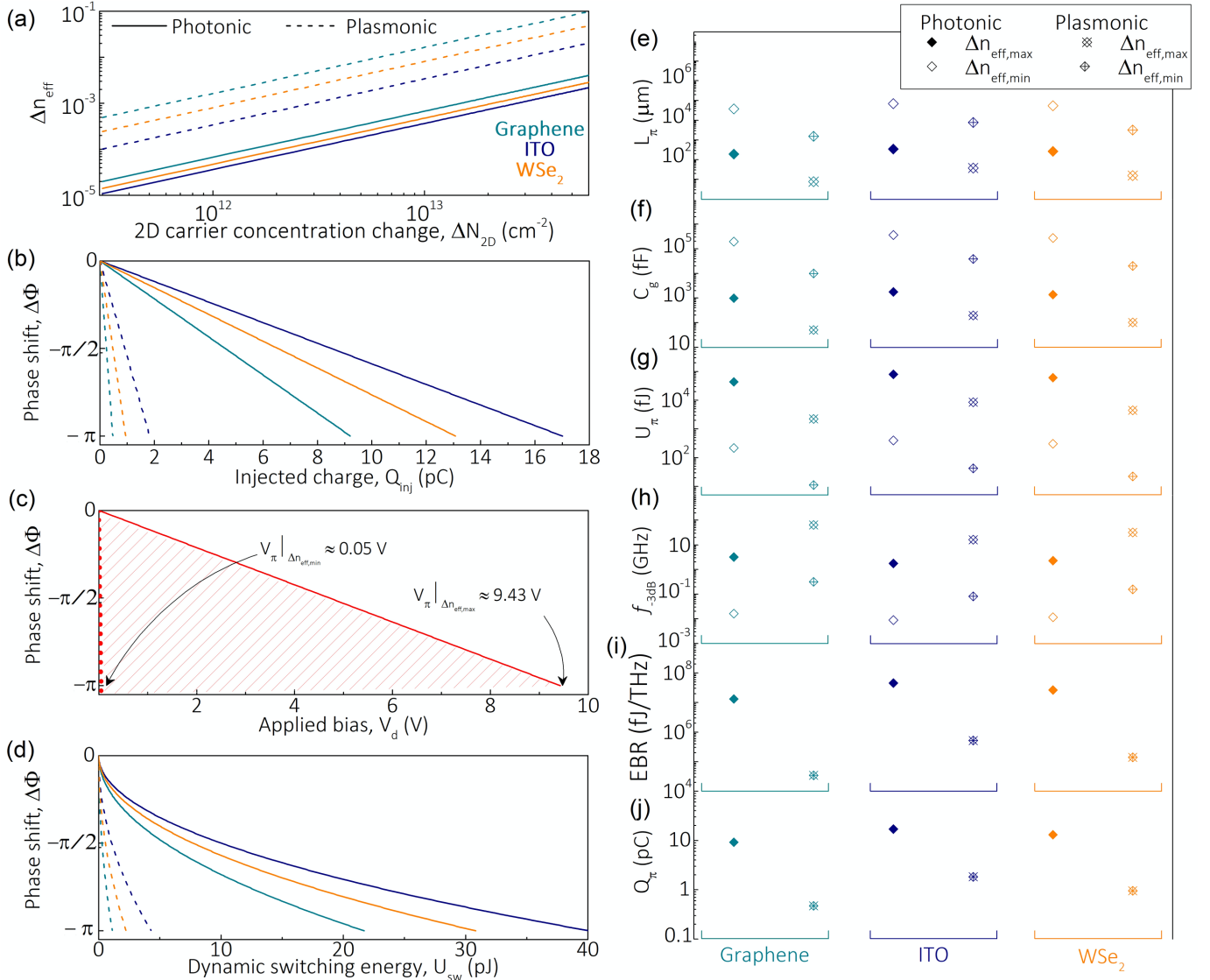

Fig. 4. Phase modulation performance analysis. (a) Effective index change, $\Delta n_{eff}$ vs. 2-dimensional carrier concentration change, $\Delta N_{2D}$ (cm$^{-2}$) for all the different modes in Fig. 3. Solid lines represent photonic modes and dashed lines represent plasmonic structures. (b) Phase shift, $\Delta\Phi$ vs. the injected (induced by the gate) charge, $Q_{inj}$ (pC); (c) Phase shift, $\Delta\Phi$ vs. drive voltage bias, $V_d$ (V); the dotted line corresponds to the minimum $\Delta n_{eff}$ sweep in (a) corresponding to the maximum length in (e) and the solid line corresponds to the maximum $\Delta n_{eff}$ sweep in (a) corresponding to the minimum length in (e). (d) Phase shift, $\Delta\Phi$ vs. the dynamic switching energy, $U_{sw}$ (pJ); and (e-j) Different performance metrics for the different phase shifters corresponding to maximum and minimum carrier concentration sweep from (a), including deivce length for $\pi$ phase shifts, $L_\pi$ (μm); gate capacitance, $C_g$ (fF); $\pi$-phase switching energy, $U_\pi$ (fJ); -3 dB cutoff speed, $f_{-3dB}$ (GHz); the energy bandwidth ratio, EBR (fJ/THz); and the charge required for $\pi$ phase shift, $Q_\pi$ (pC). Panels (b) and (d) show only results for $\Delta n_{eff,max}$ and corresponding $L_{\pi,min}$ for all schemes.

## VI. Discussion and Conclusion

Prior to concluding, it would be instructional to compare these three relatively novel phase modulation schemes with the venerable lithium niobate (LiNbO$_3$), which has recently seen a renaissance associated with development of thin film waveguides integrated on silicon [45]. Being a voltage-driven scheme, the charge in a LiNbO$_3$ modulator resides on the electrodes and the field between them changes the permittivity via an electro-optic i.e. Pockels effect. One can derive the expression for the phase change cross-section (Appendix D) as

$$\sigma_\Phi = K \frac{\pi t_a}{\lambda} \frac{e n^4 r_{33}}{\varepsilon_0 \varepsilon_{DC}} \quad (38)$$

where $n = 2.2$ is the refractive index, $r_{33} = 30$ pm/V is the Pockels coefficient, $\varepsilon_{DC} = 82$ is DC dielectric constant, $t_a$ is the active layer thickness, and $K \leq 1$ is the DC field confinement coefficient. Substituting these values, we obtain

$$\sigma_\Phi \approx 3 \times 10^{-18} cm^2 \times t_a(nm) \quad (39)$$

As one can see, in the photonic waveguides, thickness of the LiNbO$_3$ active layer is easily 100's of nanometers, bringing $\sigma_\Phi$ into $10^{-16} - 10^{-15}$ cm$^2$ range (for comparison, Table I lists the low-dimensional materials considered here). Therefore, it would be extremely difficult with photonic waveguides to compete with LiNbO$_3$ or any other traditional voltage driven modulators, such as polymers [46], GaAs [47] or InP QW's [48]. But in plasmonic waveguides, with active layers' thicknesses on the scale of ~10 nm, the current driven modulators investigated here do hold advantage over their voltage driven counterparts.

The advantage of voltage (or, better said, field–) driven over current driven modulators is easy to discern. Each carrier injected in ITO or graphene only changes the permittivity locally, while each carrier injected into an electrode together with its image on the counter electrode engenders electric field and changes the permittivity through entire active layer thickness. Obviously, in photonic waveguides, one can fill the entire 100's of nanometers thick waveguide with active material making voltage driven modulators more attractive than current driven modulators, if footprint can be ignored for the PIC design. But in plasmonic modulators, the thickness of the optical mode is small, hence the advantages of voltage driven scheme shrinks over current-driven designs, while adding an additional benefit of micrometer compactness beneficial for dense PIC layouts.

In conclusion, we have carried out a general analysis of different emerging materials for application in integrated electro-optic modulation. We have shown that performance of electro-optic modulator can be estimated on the basis of just two parameters; one is material related – phase change cross-section, $\sigma_\Phi$; and the other is the waveguide effective cross-section, $S_{eff}$. The ratio of these two cross-sections determines the number of carriers (charge) required to shift phase by 180º.

Three materials considered here – graphene, TMD and ITO, all have comparable phase change cross-sections, and it makes ITO competitive for phase modulation, which is strikingly different from electro-absorption based modulators where performance of ITO or any other free carrier modulator is orders of magnitude inferior to graphene and TMD [38]. For instance, one advantage of ITO is that the active layer can be made relatively thick which allows injection of large charge density per unit area and thus reduction of the length of the modulator, which is particularly important for plasmonic modulators where extra length leads to higher insertion loss due to absorption in the metal; i.e. the optimum length is ~1-5 µm [40]. Single layer graphene and TMD do show large phase modulation per each injected carrier, but the density of carriers that can be injected is limited by the small density of states in the monolayer. Probably multilayer graphene would be a more ideal medium for compact modulators. The same cannot be said about TMDs since multilayer TMD exhibits strong indirect absorption.

Overall one can say that as any emerging materials, monolayers of graphene and TMDs are overrated for electro-optic modulation. They do show excellent performance per each carrier, but the net effect quickly saturates as one runs out of the available densities of states in a monolayer, so it is difficult to achieve 180º phase shift within a short length. On the other hand, ITO is less glamorous in terms of performance per carrier compared to these 2D materials, but one can put more carriers into it facilitating an overall decent effect as demonstrated in [8], [9].

When compared with existing electro-optic materials (e.g. LiNbO$_3$, polymers, GaAs, InP) all of which are voltage driven, i.e. with no carrier injection, the three current (or injection) driven materials considered here are highly competitive in the thin plasmonic waveguides but much less so in the photonic waveguides. Of course, besides the fundamental performance parameters discussed here; there are many factors that may influence which material to use. These factors are mostly related to the ease of fabrication and integration on an MOS compatible Si platform and may turn out to be decisive when it comes to moving the modulators out of the lab and into commercial space. Nevertheless, we hope that the simple yet rigorous comparative analysis of fundamental performance characteristics of electro-optic modulators accomplished here, will be useful as a guide to researchers looking for the best fit to any particular application niche.

## Appendix

### A. Active Layers with Finite Thickness

When the active layer has finite thickness $t$ one can evaluate integral in (1) as

$$\int_{y_0-t_a/2}^{y_0+t_a/2} \int_{x_0-w_a/2}^{x_0+w_a/2} \delta\varepsilon_r(x,y) \mathbf{E} \cdot \mathbf{E}^* dxdy$$

$$= \delta\varepsilon_r E^2(x_0,y_0) \int_{y_0-t_a/2}^{y_0+t_a/2} \int_{x_0-w_a/2}^{x_0+w_a/2} e_a^2(x,y) dxdy \quad (A1)$$

$$= \delta\varepsilon_{2D} E^2(x_0,y_0) t_a^{-1} \int_{y_0-t_a/2}^{y_0+t_a/2} \int_{x_0-w_a/2}^{x_0+w_a/2} e_a^2(x,y) dxdy$$

where we have introduced 2D dielectric constant change as

$$\delta\varepsilon_{2D} = \delta\varepsilon_r t_a \quad (A2)$$

Therefore

$$S_{eff} = w_a t_a \iint \hat{z} \cdot (\mathbf{e} \times \mathbf{h}^*) dxdy \Big/ \int_{y_0-t_a/2}^{y_0+t_a/2} \int_{x_0-w_a/2}^{x_0+w_a/2} e_a^2(x,y) dxdy \quad (A3)$$

Obviously in the limit $t_a \to 0$ (A3) becomes (4).

*B. Graphene*

Now we can evaluate the optical conductance of graphene sheet below the absorption edge as $G = e^2/4\hbar$, so the 2D dielectric constant due to interband transitions is

$$\delta\varepsilon_{2D,ib} = iG/\varepsilon_0\omega = ie^2/4\varepsilon_0\hbar\omega = i\pi c\alpha_0/\omega \quad (A4)$$

Substitution into (5) yields for the interband absorption coefficient of undoped graphene sheet placed into the waveguide as

$$\alpha_{ib0} = 2\,\text{Im}(\delta\beta) = \pi\alpha_0 \frac{w_a}{n_{eff} S_{eff}} \quad (A5)$$

When the graphene is doped and the absorption edge is at photon energy $2E_F$, this absorption coefficient must be multiplied by the probability of having empty states in the conduction band at energy $\hbar\omega/2$,

$$F = \left( \exp\frac{E_F - \hbar\omega/2}{kT} + 1 \right)^{-1} \quad (A6)$$

which leads to Eq.(7). Now, for the intraband (free carrier) absorption – the conductivity is

$$G_{fc} = \frac{e^2(\gamma + j\omega)}{\pi\hbar^2(\omega^2 + \gamma^2)} E_F \quad (A7)$$

and

$$\delta\varepsilon_{2D,fc} = iG_{fc}/\varepsilon_0\omega = \frac{e^2 E_F}{\pi\varepsilon_0\hbar^2\omega(\omega^2+\gamma^2)}(i\gamma - \omega)$$

$$= \frac{4c\alpha_0}{\omega}(i\gamma - \omega)\frac{E_F}{\hbar(\omega^2+\gamma^2)} \quad (A8)$$

Substitution into (5) yields for the free carrier absorption coefficient

$$\alpha_{fc} = 2\,\text{Im}(\delta\beta) = \frac{w_a}{n_{eff} S_{eff}} \alpha_0 \frac{4\gamma E_F}{\hbar(\omega^2+\gamma^2)} \quad (A9)$$

*C. Indium Tin Oxide (ITO)*

For ITO – two dimensional dielectric constants is

$$\varepsilon_{2D}(\omega) = \varepsilon_\infty t_a - \frac{e^2 n_{2D}/\varepsilon_0 m^*}{\omega^2+\gamma^2} + i\frac{e^2 n_{2D}/\varepsilon_0 m^*}{\omega^2+\gamma^2}\frac{\gamma}{\omega} \quad (A10)$$

where $t_a$ is the active layer thickness, $m^*$ is optical effective mass, and $\varepsilon_\infty$ is the 3D dielectric constant due to valence electrons. The change of the real part dielectric constant is then

$$\delta\varepsilon_{2D}(\omega) = -\frac{e^2/\varepsilon_0 m^*}{\omega^2+\gamma^2}\delta n_{2D} = -4\pi\frac{\hbar}{m^*(\omega^2+\gamma^2)}c\alpha_0\delta n_{2D} \quad (A11)$$

So the change of the real part of the propagation constant according to (5) is then

$$\text{Re}(\delta\beta) = \frac{\omega}{2n_{eff}c}\frac{\delta\varepsilon_{2D} w_a}{S_{eff}} \equiv \delta n_{eff}\frac{\omega}{c} \quad (A12)$$

and

$$\delta n_{eff} = -\frac{w_a}{2n_{eff} S_{eff}}\frac{e^2/\varepsilon_0 m^*}{\omega^2+\gamma^2}\delta n_{2D} \quad (A13)$$

*D. Phase Shift Cross Section for Lithium Niobate or Any Other Voltage Driven Scheme.*

In a voltage driven scheme, the dielectric permittivity is modulated by the electric field while the charge is stored not in the active layer itself, but on two electrodes encasing the charge. The induced electric field is related to the 2D electron density on electrodes as

$$E = Ke\delta n_{2D}/\varepsilon_0\varepsilon_{DC} \quad (A14)$$

where $\varepsilon_{DC}$ is the DC dielectric constant and K < 1 is the coefficient describing the fact that the field is not entirely confined inside the active layer. For a planar parallel capacitor K = 1. Change of the 2D optical dielectric constant can be found as

$$\delta\varepsilon_{2D} = t_a n^4 r_{33} E \quad (A15)$$

where $n$ is the refractive index and $r_{33}$ = 30 pm/V is the electro-optic coefficient. Next, we obtain

$$\Delta\Phi = \delta\beta L = \frac{\omega}{2n_{eff}c}\frac{w_a L}{S_{eff}}\frac{n^4 r_{33} Ke}{\varepsilon_0\varepsilon_{DC}}\delta n_{2D} = \frac{\delta Q}{en_{eff} S_{eff}}\sigma_\Phi \quad (A16)$$

where the phase change cross section is

$$\sigma_\Phi = \frac{\pi t_a}{\lambda}\frac{en^4 rK}{\varepsilon_0\varepsilon_{DC}} \quad (A17)$$

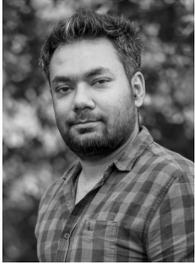
**Rubab Amin** received his B.S. degree with the distinction *Summa Cum Laude* in Electronics and Telecommunication Engineering from North South University, Dhaka, Bangladesh in 2011. He received his PhD in Electrical Engineering from The George Washington University, Washington DC, USA in 2020. Currently, he is working as a postdoctoral scientist in the OPEN Lab team at The George Washington University, Washington DC. His research is focused on theory, design and demonstration of novel nanoscale electro-optic integrated modulators on Si-photonic platforms. His research interests include nanophotonics, plasmonics, electro-optic modulators, transparent conducting oxides and low-dimensional material photonics. He has served as the vice-president of the OSA/SPIE student chapter at GWU for multiple terms. He received several awards including the SPIE optics and photonics education scholarship, NSF I-Corps award, GW research days and R&D showcase best poster awards, university fellowship, and international amigo scholarship.

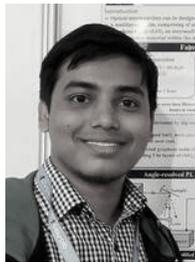
**Rishi Maiti** (M'19) received his M.Sc. degree in Physics from IIT Kharagpur, India in 2012, and then received his Ph.D. degree in 2017. His PhD research topic was studies on hybrid graphene nanostructures for optoelectronic devices. He joined the University of Brescia, Italy as a visiting scholar under Erasmus Mundus scholarship in 2016. Currently, he is a Post-Doctoral Fellow in the George Washington University. His research interests include the nanophotonic devices, Electro-optic modulator, optical interconnect, novel materials, plasmonics & metamaterials, Tunnel junction & Smart window, as well as their promising applications towards fully integrated photonic circuit.

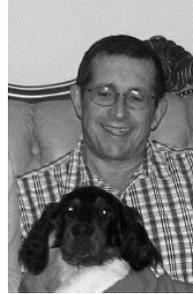
**Jacob B. Khurgin** received the Ph.D. degree from the Polytechnic University of New York, New York City, New York, USA. He has been a Professor of Electrical and Computer Engineering, Johns Hopkins University (JHU), Baltimore, MD, USA, since 1988. Prior to that, he was a Senior Member of Research Staff with Philips NV where he developed various display components and systems including 3-D projection TV and visible lasers pumped by electron beam. In his 30 years with JHU, he had made important contributions in the fields of nonlinear optics, semiconductor optoelectronic devices, quantum-cascade lasers, optical communications, terahertz technology, slow light, plasmonics, opto-mechanics, and fundamental condensed matter physics. He had authored more than 300 technical papers, 500 conference presentations, five book chapters, and held 40 patents. His research interests include optical and electronic solid state devices. Prof. Khurgin is a Fellow of American Physical Society and Optical Society of America.

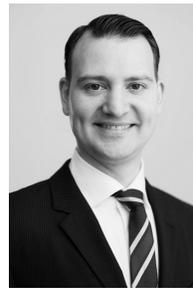
**Volker J. Sorger** (M'07) Volker J. Sorger is an Associate Professor in the Department of Electrical and Computer Engineering and the director of the Orthogonal Physics Enabled Nanophotonics (OPEN) Lab at the George Washington University. His research areas include opto-electronics, nanophotonics, intelligent information processing, and photonic AI systems. Among his breakthroughs are the first micrometer-compact GHz modulator, strainoptronics photodetector, plasmon laser, and innovations such as multilevel photonic memory and photonic tensor core processor. For his work, Dr. Sorger received multiple awards among are the Presidential Early Career Award for Scientists and Engineers (PECASE), the AFOSR Young Investigator Award (YIP), the Hegarty Innovation Prize, and the National Academy of Sciences award of the year. Dr. Sorger is the editor-in-chief of NANOPHOTONICS, holds the position of the OSA Division Chair for 'Photonics and Opto-electronics' serving at the boards of OSA & SPIE. He is a senior member of IEEE, OSA & SPIE, and the founder of Optelligence LLC.